\documentclass[a4paper]{article} 
\usepackage{latexsym,amssymb,amsmath} 
  
\begin{document} 
\title{Graviton oscillations in the two-brane world} 
\author{Andrei O.\ Barvinsky\thanks{barvin@td.lpi.ru}\\ 
Theory Department, Lebedev Physics Institute,\\
Leninsky Pr.\ 53,
Moscow 117924, 
Russia.\\[3mm]
Alexander Yu.\ Kamenshchik\thanks{sasha.kamenshchik@centrovolta.it}\\ 
L.\ D.\ Landau Institute for Theoretical Physics\\ 
of Russian 
Academy of Sciences,\\ Kosygina str.\ 2, Moscow 117334, Russia,\\ 
and\\ 
Landau Network --- Centro Volta,\\ 
Villa Olmo, via Cantoni 1, 22100 Como, 
Italy.\\[3mm]
Claus Kiefer\thanks{kiefer@thp.uni-koeln.de}\\
Institut f\"ur Theoretische Physik, Universit\"at zu K\"oln,\\  
Z\"ulpicher Str.\ 77, 50937 K\"oln, Germany.\\[3mm] 
Andreas Rathke\thanks{ra@thp.uni-koeln.de}\\
Physikalisches Institut, Universit\"at Freiburg,\\ 
Hermann-Herder-Str.\ 3, 79104 Freiburg, Germany,\\ 
 and\\ 
Institut f\"ur Theoretische Physik, Universit\"at zu K\"oln,\\  
 Z\"ulpicher Str.\ 77, 50937 K\"oln, Germany.} 
\date{} 
\maketitle 
\begin{abstract} 
We study the braneworld effective action in the two-brane Randall-Sundrum
model.  In the framework of this essentially-nonlocal action we reveal the
origin of an infinite sequence of gravitational wave modes --- the usual
massless one as well as the tower of Kaluza-Klein massive ones. Mixing of
the modes, which parametrically depends on the background value of the modulus
of the extra dimension, can be interpreted as radion-induced gravitational-wave
oscillations, a classical analogue to meson and neutrino oscillations. We
show that these oscillations arising in M-theory-inspired braneworld setups
could lead to effects detectable by gravitational-wave interferometers.
\end{abstract}

In recent years the old Kaluza-Klein idea of a world with additional
dimensions has received new impacts from string-inspired attempts to resolve
the hierarchy problem \cite{ADD} (see \cite{kubyshinrubakov} for reviews).
Especially fruitful is the idea of implementing warped manifolds, for which
the metric of the four-dimensional space-time can depend on the additional
dimensions.  The most popular models using warped geometries are the
Randall-Sundrum (RS) models \cite{RS1,RS2}, where our observable Universe is
considered as a four-dimensional brane embedded into a five-dimensional
anti-De~Sitter (AdS) bulk space.
 
The particular properties of the AdS geometry combined with the requirement of 
a ${\mathbb Z}_2$-orbifold symmetry allow to describe the trapping of the 
graviton zero-mode 
on the four-dimensional brane \cite{RS2}. 
Displacing the second four-dimensional brane in the bulk \cite{RS1}, one can 
propose a resolution to the hierarchy problem.  However, the two-brane models 
have some important additional features.   
The most interesting effects are the existence of massive graviton modes 
and the nontrivial dependence of 
the mixing between different graviton states on the inter-brane distance 
called the radion field.  
 
In the following we assume that we live on the positive-tension 
brane, which is motivated by considering the RS setup as a simplified model 
for M-theory constructions, in which the standard-model fields usually reside 
on this brane (cf.\ \cite{Pyr}). The effects seen by an observer on the 
negative-tension brane will be analyzed in \cite{tocome}. 
 
In our preceding paper \cite{we} we have studied the four-dimensional
effective action in the two-brane model.  This action is a functional of two
induced metrics and linearized radion fields on the two four-dimensional
branes.  Using the results of \cite{we}, we shall study here the effects of
radion-induced graviton oscillations in the two-brane world and investigate
their phenomenological consequences.  For this purpose it will be sufficient
to consider the sector of transverse-traceless modes, i.\,e. ``graviton''
modes, without making reference to the conformal modes of the metrics and the
radions because these do not couple to the transverse-traceless modes at the
linearized level. 
Nevertheless, a parametrical influence of the radion in its full
nonlinear form on the transverse-traceless modes persists due to the
dependence of the transverse-traceless sector of the action on the warp factor
which is nothing but the non-perturbative generalization of the linearized
radion. This dependence will turn out to have interesting phenomenological
implications.

In the RS setup \cite{RS1}, the braneworld effective action is induced from 
the five-dimensional bulk spacetime as follows. We start with the theory 
having the action of the five-dimensional gravitational field with metric 
$G=G_{AB}(x,y)$, $A=(\mu,5),\,\mu=0,1,2,3$, propagating in the bulk spacetime 
$x^A=(x,y),\,x=x^\mu,\,x^5=y$, and matter fields $\phi$ confined to the two 
branes $\Sigma_\pm$, 
\begin{equation}
S[\,G,g,\phi\,]=S_5[\,G\,] + S_4[\,G,g,\phi\,], \label{action}
\end{equation}
where the bulk-space part of the action is given by
\begin{equation}
     S_5[\,G\,]=\frac1{16\pi G_5} 
     \int\limits_{M^5} d^5x\,G^{1/2} 
     \left(\,\vphantom{I}^5\!R(G)-2\Lambda_5\right),  \label{action1} 
\end{equation} 
and the brane-part of the action is
      \begin{equation} 
     S_4[\,G,g,\phi\,]=
     \sum\limits_{\pm}\int_{\Sigma_\pm}\! 
     d^4x\,\Big( L_m(\phi,\partial\phi,g)
     -g^{1/2}\sigma_\pm
     + \frac1{8\pi G_5}[K]\Big). \label{actionbr}
     \end{equation}

     The branes are enumerated by the index $\pm$ and carry induced metrics
     $g=g_{\mu\nu}(x)$ and matter-field Lagrangians
     $L_m(\phi,\partial\phi,g)$. The bulk part of the action (\ref{action1})
     is characterized by the five-dimensional gravitational constant $G_5$ and
     the cosmological constant $\Lambda_5$, while the brane parts
     (\ref{actionbr}) have four-dimensional cosmological constants
     $\sigma_\pm$.  The bulk cosmological constant $\Lambda_5$ is negative
     and, therefore, is capable of generating an AdS geometry, while the brane
     cosmological constants play the role of brane tensions and, depending on
     the model, can be of either sign. The Einstein-Hilbert bulk action
     (\ref{action1}) is accompanied by terms in (\ref{actionbr}) containing
     the jumps of extrinsic curvatures traces $[K]$ associated with both sides
     of each brane \cite{ChR}.  Solving the five-dimensional Einstein
     equations with prescribed values of the four-dimensional metric on the
     branes, we have obtained in \cite{we} the tree-level effective
     four-dimensional action.
 
In the RS two-brane setup, the fifth dimension has the topology of a circle
labeled by the coordinate $y$, $-d\leq y\leq d$, with an orbifold ${\mathbb
  Z}_2$-identification of points $y$ and $-y$. The branes are located at
antipodal fixed points of the orbifold, $y=y_\pm,\,y_+=0,\,|y_-|=d$. They are
empty, i.\,e.\ $L_m(\phi,\partial\phi,g_{\mu\nu})=0$, and their tensions are
opposite in sign and fine-tuned to the values of $\Lambda_5$ and $G_5$,
     \begin{eqnarray} 
     &&\Lambda_5=-6/l^2,\quad\quad 
     \sigma_+=-\sigma_-=3/(4\pi G_5l).   \label{RS1} 
     \end{eqnarray} 
Then this model admits a solution to the Einstein equations with an AdS 
metric in the bulk ($l$ is its curvature radius), 
     \begin{eqnarray} 
     ds^2=dy^2+e^{-2|y|/l}\eta_{\mu\nu}dx^\mu dx^\nu,  \label{RS2} 
     \end{eqnarray} 
$0=y_+\leq|y|\leq y_-=d$, and with a flat induced metric 
$\eta_{\mu\nu}$ on both branes \cite{RS1}. With the fine tuning 
(\ref{RS1}) this solution exists for arbitrary brane separation 
$d$. 
 
Now consider small metric perturbations $\gamma_{AB}(x,y)$ on the  
background of this solution \cite{RS2,GT,GKRChGR}, 
     \begin{eqnarray} 
     ds^2=dy^2+e^{-2|y|/l}\eta_{\mu\nu}dx^\mu dx^\nu 
     +\gamma_{AB}(x,y)\,dx^Adx^B.                   \label{metric} 
     \end{eqnarray} 
Then this five-dimensional metric {\em induces} on the branes two 
four-dimensional metrics of the form 
    \begin{eqnarray} 
    g^\pm_{\mu\nu}(x)= 
    a^2_\pm\,\eta_{\mu\nu}+\gamma^\pm_{\mu\nu}(x),  \label{metric1} 
    \end{eqnarray} 
where the warp factors $a_\pm=a(y_\pm)$ are given by 
    $a_+=1$ and  
$a_-=e^{-d/l}\equiv a$,  
and $\gamma^\pm_{\mu\nu}(x)$ are the perturbations by which the brane 
metrics $g^\pm_{\mu\nu}(x)$ differ from the (conformally) flat 
metrics of the RS solution (\ref{RS2}). The variable $a$ represents 
the modulus --- the global part of the radion field determining the 
interbrane separation.

In \cite{we} we have derived the effective braneworld action in 
terms of the four-dimensional on-brane metrics (\ref{metric1})  
and non-local matrix-valued form 
factors. We call the part of the action quadratic in $h^\pm_{\mu\nu}(x)$ --- 
the transverse-traceless parts of the full metric perturbations  
$\gamma_{\mu\nu}^\pm(x)$ on the branes --- the graviton sector.  
It reads as the following $2\times2$ quadratic form, 
    \begin{equation}  
    S_{\rm grav}[\,h^\pm_{\mu\nu}\,]  
    =\frac1{16\pi G_4}\int  
    d^4x\,\frac12\,  
    {\bf h}^T  
    \,\frac{{\bf  F}(\Box)}{l^2}\,  
    {\bf h},        \label{4.24}  
    \end{equation} 
in terms of columns of the metric perturbations  
    \begin{equation}  
    {\bf h}=  
    \left[\begin{array}{c}  
    \,\,h^+ \\  
    \,\,h^-\,  
    \end{array}\right]       
\end{equation} 
(${}^T$ denotes their transposition into rows and we omit from now on the
tensor indices) and the special nonlocal operator ${\bf F}(\Box)$ (we also use
$G_4 \equiv G_5/l$). As was shown in Ref.\ \cite{we}, the operator ${\bf
  F}(\Box)$ is a complicated non-linear function of the D'Alembert operator
$\Box$, expressed by means of Bessel and Neumann functions of arguments $l
\sqrt{\Box}$ and $l\sqrt{\Box}/a$. In this paper, we study in detail its
properties in the \emph{low-energy limit}, when $l \sqrt{\Box} \ll 1$ but when
$l \sqrt{\Box}/a$ can take arbitrary values in view of the smallness of the
parameter $a=e^{-d/l}$ (large interbrane distances).  In this limit the kernel
of the action ${\bf F}(\Box)$ reads
    \begin{equation}  
    {\bf F}(\Box) \approx \frac{l^2\Box}{2} 
    \left[\begin{array}{cr}  
    1   
    &1/J_2[l\sqrt\Box/a]\\[3mm]   
    1/J_2[l\sqrt\Box/a]&  
    -\frac 2{l\sqrt{\Box} a}  
    \frac{J_1[l\sqrt\Box/a]}{J_2[l\sqrt\Box/a]}\,  
    \end{array}\right].      \label{7.4.3}  
    \end{equation} 
  
A typical way to extract the particle content from an action with a 
matrix-valued kernel is its diagonalization in terms of normal modes.  
However, in view of the nonlocality of ${\bf F(\Box)}$ the number of 
propagating modes enormously exceeds the number of entries in the 
$2\times2$-matrix ${\bf F}(\Box)$ and they do not diagonalize the 
quadratic action (\ref{4.24}) in the usual sense. The propagating modes 
${\bf h}_i(x)=h_i(x){\bf v}_i$ are the zero modes of ${\bf F}(\Box)$ 
which solve the matrix-valued nonlocal equation  
${\bf F}(\Box){\bf h}_i(x)=0$. The consistency of the latter,  
$\det{\bf F}(\Box)=0$, yields a mass spectrum of the theory given by 
the roots of this equation, i.\,e.\ $\Box=m_i^2$, so that the $h_i(x)$ above 
are massive Klein-Gordon modes, $(\Box-m_i^2)\,h_i(x)=0$, and the isotopic 
vectors of the propagating modes ${\bf v}_i$ are zero eigenvectors 
of ${\bf F}(m_i^2)$,
${\bf F}(m_i^2){\bf v}_i=0$. This gives
 the Kaluza-Klein 
spectrum which contains the massless mode $i=0$, $m_0=0$, and the 
tower of massive modes with masses $m_i=a j_i/l$ given in the low-energy  
approximation of (\ref{7.4.3}), $a\ll1$, by the roots of the first order  
Bessel function, $J_1(j_i)=0$. The isotopic structure of their  
${\bf v}_i$ is  
    \begin{equation}  
    {\bf v}_0= \frac{\sqrt 2}l 
    \left[\begin{array}{c}  
    \,1 \\  
    \,\,a^2\,  
    \end{array}\right],\,\,\, 
    {\bf v}_i= \frac{{\sqrt 2}a}l 
    \left[\begin{array}{c}  
    1/J_2(j_i) \\  
    -1\,  
    \end{array}\right]\ .  \label{v} 
    \end{equation} 
The action (\ref{4.24}) is not, however, diagonalizable in the basis of these 
states because under the decomposition ${\bf h}(x)=\sum_{i=0}h_i(x){\bf v}_i$ 
(with off-shell coefficients $h_i(x)$) the cross terms intertwining 
different $i$-s are nonvanishing, ${\bf v}_i^T {\bf F}(\Box) {\bf v}_j\neq 0$. 
 
A crucial observation is, however, that the diagonal and nondiagonal 
terms of this expansion are linear and bilinear, respectively, in on-shell  
operators $\Box-m_i^2$,  
\begin{align} 
&{\bf v}^T_i {\bf F}(\Box) {\bf v}_i  
= \Big({\bf v}^T_i \frac{d{\bf F}(\Box)}{d\Box} 
{\bf v}_i\Big)_{\Box = m_i^2} (\Box - m_i^2),\label{99}\\ 
&{\bf v}^T_i {\bf F}(\Box) {\bf v}_j  
= M_{ij}(\Box)(\Box - m_i^2)(\Box - m_j^2),\,\,i\neq j, \label{100} 
\end{align} 
where higher powers of $(\Box-m_i^2)$ have been dropped in (\ref{99}) and
$M_{ij}(\Box)$ is nonvanishing at both $\Box=m_i^2$ and $\Box=m_j^2$.
Therefore, the nondiagonal terms of the action do not contribute to the
residues of the Green's function ${\bf G}(\Box)$ of ${\bf F}(\Box)$, which
turns out to be given by direct products of the ${\bf v}_i$ and their
transposes,
\begin{equation}  
{\bf G}(\Box) = \sum_{i=0}  
\frac{{\bf v}_i{\bf v}_i^T} 
{\Box - m_i^2},               \label{Green} 
\end{equation} 
the normalization of ${\bf v}_i$ being 
chosen in such a way as to render a unit 
coefficient of $(\Box-m_i^2)$ in (\ref{99}). 
 
In passing, we 
note that the problem of non-orthogonal physical modes ${\bf v}_i$ is well
known in the theory of atomic and nuclear resonances \cite{More}, where these
modes are called Siegert states as opposed to the so-called Kapur-Peierls
states, ${\bf b}_s(\Box)$, which diagonalize the {\em matrix} Hamiltonian (the
analogue of our ${\bf F}(\Box)$), ${\bf F}(\Box) {\bf b}_s(\Box) =
\lambda_s(\Box) {\bf b}_s(\Box)$, $s=1,2$, ${\bf b}^T_s(\Box){\bf
  b}_{s'}(\Box)=\delta_{ss'}$. In this formalism the Siegert states and their
masses (energy levels) arise as zeroes of the eigenvalues,
$\lambda_s(m_i^2)=0$, and turn out to be related to the Kapur-Peierls states,
${\bf v}_0\sim {\bf b}_1(m_0^2)$, ${\bf v}_i\sim {\bf b}_2(m_i^2)$, $i\neq 0$,
up to some nontrivial normalization coefficients. The details of this
formalism will be presented in 
\cite{tocome}.
 
The above method provides us with the conventional  
particle interpretation of the propagating modes of the nonlocal operator 
${\bf F}(\Box)$ and their role in its Green's function (\ref{Green}) 
mediating the gravitational effect of matter sources. Amended by the  
matter action on the branes (cf.\ \cite{we}) the effective action of the  
graviton sector reads 
\begin{align} 
  S\,[\,h^\pm_{\mu\nu}\,]  
    =\int  
    d^4x\,\left(\frac1{32\pi G_4}  
    {\bf h}^T  
    \,\frac{{\bf  F}(\Box)}{l^2}\,  
    {\bf h}+  \frac12{\bf h}^T {\bf T} \right), 
\end{align} 
where ${\bf T}$ is the column vector of the stress-energy 
tensors on the branes.  Varying this action with respect to ${\bf h}$ we 
obtain the linearized equations of motion, their solution   
${\bf h} =-8\pi G_4 l^2{\bf G}_\text{ret} {\bf T}$ being expressed 
in terms of the retarded version of the Green's function (\ref{Green}).   
 
We shall now restrict ourselves to observers living on the 
$\Sigma_+$-brane (``visible brane'').  For simplicity we also restrict our 
attention to frequencies below the mass threshold of the second massive mode 
$m_2$. Then using the spectral representation (\ref{Green}) and the 
structure of ${\bf v}_{0,1}$ given by (\ref{v}) we find for $h^+$,  
      \begin{equation} 
       h^+= -\left.\frac{16\pi G_4}{\Box} 
       \right|_{\rm ret} 
(T^++a^2 T^-) 
        -\left.\frac{16\pi G_4}{\Box-m_1^2} 
        \right|_{\rm ret} 
\left(\frac{a^2}{\mathcal{J}_2^2}T^+- 
       \frac{a^2}{\mathcal{J}_2}T^-\right), 
\end{equation} 
where $\mathcal{J}_2 \equiv J_2[lm_1/a] \approx 0.403$.  
We consider astrophysical sources at ${\mathbf x}=0$ on both branes of equal  
intensity with a harmonic time dependence  
$T^\pm(t,{\mathbf x})=\mu e^{-i\omega t}\delta({\mathbf x})$.  If the 
frequency of the source is above the mass threshold of the massive mode, 
i.\,e.\ $\omega >m_1$, both modes, the massless and the first  
massive one, are excited and produce long range gravitational waves.  
At distance $r$ from the source the waves on  
each brane are given by a mixture of massless and massive  
spherical waves.  On the 
$\Sigma_+$-brane this superposition is given by the sum of the  
contributions  
 \begin{flalign} 
 h^+[\,T^+] & = A\, e^{-i\omega t} 
 \left(e^{i\omega r} +  \frac{a^2}{\mathcal{J}_2^2}  
 e^{i \sqrt{\omega^2-m_1^2}r} 
                              \right)\ ,       \label{h+of+} \\  
 h^+[\,T^-] & = A a^2\, e^{-i\omega t}  
 \left( e^{i\omega r} - \frac{1}{\mathcal{J}_2}  
 e^{i \sqrt{\omega^2-m_1^2}r} \right)\ ,          \label{h+of-} 
 \end{flalign} 
of sources on the $\Sigma_+$- and $\Sigma_-$-brane, respectively, where  
$A=4\,G_4\,\mu/r$ is the amplitude of the massless mode produced on  
the $\Sigma_+$-brane.  The 
amplitudes detected by a gravitational-wave interferometer are given by the 
absolute values of (\ref{h+of+}) and (\ref{h+of-}), 
\begin{align} 
\big|h^+[\,T^+]\big| &= \mathcal{A}^+ 
\left[ 1- \frac{4a^2 \mathcal{J}^2_2}{(\mathcal{J}^2_2+a^2)^2}  
\sin^2 \left( \frac{\pi r}{L}\right) 
\right]^{1/2}, 
\label{absh+of+} \\ 
\big|h^+[\,T^-]\big| &=  
\mathcal{A}^- 
\left[ 1 + \frac{4\mathcal{J}_2}{(\mathcal{J}_2-1)^2}  
\sin^2 \left( \frac{\pi r}{L}\right) 
\right]^{1/2}. 
\label{absh+of-} 
\end{align} 
Here $L$ is the oscillation length of the amplitude modulation of the 
gravitational wave (GW), 
\begin{equation} 
L= 2\pi \left(\omega - \sqrt{\omega ^2-m_1^2} \right)^{-1}\simeq  
\frac{2\pi}{m_1} ,\label{length} 
\end{equation} 
where the approximation corresponds to $m_1 \lesssim \omega$.  The pre-factors
of the amplitudes (\ref{absh+of+}) and (\ref{absh+of-}) are given by
\begin{align} 
\mathcal{A}^+ &= \left(1+ a^2/\mathcal{J}^2_2\right) A \approx A,
\label{prefacA+} \\ 
\mathcal{A}^- &= \left(1/\mathcal{J}_2 - 1\right) a^2 \,A 
\approx  1.5 \, a^2A, 
\label{prefacA-} 
\end{align} 
where the approximations are valid in the limit $a\ll 1$. We find oscillations 
in the amplitudes of the waves from both sources. For a GW produced by $T^+$, 
Eq.~(\ref{absh+of+}), the oscillation is suppressed 
by a factor of $a^2$ compared 
to the constant part of the amplitude in the limit of large brane separation, 
$a \ll 1$. The amplitude of the GW produced by $T^-$ (\ref{absh+of-}) is 
totally oscillating, regardless of the inter-brane distance. 
 
We can express the oscillation length of 
these radion-induced gravitational wave oscillations (RIGO's) through the AdS 
radius $l$ and the warp factor $a$, 
\begin{equation} 
L = \frac{2 \pi}{j_1} \, \frac la
\approx 1.6\, \frac la\ , 
\end{equation}
where $j_1 \approx 3,831$ ist the first root of $J_1$.
The oscillation length is inverse proportional to $a$. Graviton oscillations 
become observable when the oscillation length is of the same size as the 
arm length of a GW detector. For the ground-based interferometric detectors this 
requirement corresponds to $L \sim 10^3\,{\rm m}$.  Combining this with the 
constraint on the maximal AdS radius $l$ from sub-millimeter test 
of gravity $l 
\lesssim 10^{-4}\,{\rm m}$ \cite{Gundlach}, we find an upper limit on the warp 
factor $a \lesssim 10^{-7}$ for the oscillation length to be detectable. 
Inserting this into the ratio of the amplitudes (\ref{prefacA+}) and
(\ref{prefacA-}) we find 
\begin{equation} 
\mathcal{A}^- / \mathcal{A}^+  \lesssim 10^{-14}. 
\label{damping}
\end{equation} 
Therefore, the amplitude of a wave originating from a source on the
(``hidden'') $\Sigma_-$-brane with oscillations which are sufficiently long
to be detectable is strongly suppressed compared to a GW stemming from a
source on the $\Sigma_+$-brane itself.  A strongly oscillating wave has to be
generated by a source 14 orders of magnitude stronger than that of a weakly
oscillating one in order to be of the same magnitude, which at first sight
makes the detection of RIGO's impossible. 

However, this requirement may actually be fulfilled in brane-world setups
motivated by M-theory compactifications.  In these, the Calabi-Yau 
volume $V$
is smaller at the position of the hidden brane than at the visible brane
\cite{Witten}. 
For example, in the M-theory solution of \cite{LOSW}
the relation between $V$ and the warp factor is given by
$V(y) \sim a^6(y)$.
The vacuum expectation value (vev) of
the gaugino condensate on the hidden brane, $\eta$, depends on 
$V(y_-)$ as \cite{gaugino}
$\eta \sim V^{-9/2} \exp (- V \mathcal{S})$,
where $\mathcal{S}$ is a positive function of the Calabi-Yau moduli, 
the unified gauge coupling and its renormalization-group $\beta$-coefficient.
Therefore, $\eta$ may become very large when $V(y_-)$ becomes small. 
On the other hand, the amplitude of GW's radiated by cosmic strings is
proportional to the vev squared of the associated symmetry breaking
because the quadrupole moment of a cosmic string is given by
$\mu = \Gamma \eta^2 / \omega^{3}$,
where $\Gamma \approx 50 \ldots 100$ is a numerical coefficient 
depending on the trajectory and shape of the string loop, 
and $\omega$ is the characteristic
frequency of string oscillations \cite{VilShel}.
Thus the large gaugino vev can lead to the production of
strong-amplitude GW's on the hidden brane, easily compensating the damping
factor (\ref{damping}) \cite{tocome}.

RIGO's are in principle a generic feature of every higher-dimensional
spacetime model as there will always occur amplitude modulations in GW's,
which are a mixture of a massless mode and KK modes. In particular, their
occurrence is not affected by modulus stabilization. However, in traditional
models with flat extra dimensions, the mass of the first KK mode is so big
that it will neither be produced by astrophysical sources nor lead to
oscillation lengths of macroscopic size. In contrast to this, warped
geometries allow KK-mode masses which are so low that they can lead to
oscillations of detectable length. 
In particular, waves from sources on the
hidden brane show strong oscillations on the visible brane.
Even without assuming particularly strong sources on the hidden brane it
is conceivable to find set\-ups which combine the requirements of small KK
masses and strong amplitude modulations of GW's.  One of such set\-ups is the
Kogan-Mouslopoulos-Papazoglou version \cite{KMP} of the Karch-Randall model
\cite{KR} with two AdS$_4$ branes embedded in AdS$_5$.  In this model the
massless graviton and the lightest massive graviton are coupled to matter with
nearly equal strength and, therefore, produce strong oscillations, which are,
unfortunately, not observable because of the big oscillation length exceeding
the horizon of the AdS brane. In summary, RIGO's could be an effect which is
suitable for verifying the existence of extra dimensions in the paradigm of
warped braneworlds. 

\medskip 

A.O.B. and A.Yu.K. are grateful for the hospitality of the Theoretical Physics
Institute, University of Cologne, where a major part of this work has been
done due to the support of the DFG grant 436 RUS 113/333/0-2. A.O.B. is also
grateful to the University of Munich, where the final stage of this work was
done under the grant SFB 375.  The work of A.O.B.\ was also supported by the
Russian Foundation for Basic Research under the grant No 02-02-17054 and by
the grant of Leading Scientific Schools 1578.2003.2 , while A.Yu.K.\ was
supported by the RFBR grant No 02-02-16817 and by the scientific school grant
No 2338.2003.2 of the Russian Ministry of Industry, Science and Technology.
A.R.\ is supported by the DFG Graduiertenkolleg ``Nonlinear Differential
Equations''.

\end{document}